\documentclass[a4paper,12pt]{article}
\usepackage[pctex32]{graphics}

\begin{document}
\topmargin 0pt \oddsidemargin 0mm

\renewcommand{\thefootnote}{\fnsymbol{footnote}}
\begin{titlepage}
\begin{flushright}
INJE-TP-05-03\\
hep-th/0502128
\end{flushright}

\vspace{5mm}
\begin{center}
{\Large \bf Sourced Friedmann equations with holographic energy
density} \vspace{12mm}

{\large  Yun Soo Myung\footnote{e-mail
 address: ysmyung@physics.inje.ac.kr}}
 \\
\vspace{10mm} {\em  Relativity Research Center, Inje University,
Gimhae 621-749, Korea \\
and  Institute of Theoretical Science, University of Oregon,
Eugene, OR 97403-5203, USA}

\end{center}

\vspace{5mm} \centerline{{\bf{Abstract}}}
 \vspace{5mm}
We reexamine  cosmological applications of the holographic energy
density in the framework of sourced Friedmann equations. This
framework is suitable  because it can accommodate a macroscopic
interaction between holographic and ordinary matter naturally. In
the case that the holographic energy density decays into dust
matter, we propose a microscopic mechanism to generate an
accelerating phase. Actually, the cosmic anti-friction arisen from
the decay process induces acceleration. For examples, we introduce
two IR cutoffs of Hubble horizon and future event horizon to test
this framework. As a result, it is shown that the equations of
state for the holographic energy density are determined to be  the
same negative constants  as those for the dust matter.
\end{titlepage}
\newpage
\renewcommand{\thefootnote}{\arabic{footnote}}
\setcounter{footnote}{0} \setcounter{page}{2}

\section{Introduction}
Supernova (SUN Ia) observations suggest that our universe is
accelerating and the dark energy contributes $\Omega_{\rm
DE}\simeq 2/3$ to the critical density of the present
universe~\cite{SN}. Also  cosmic microwave background
observations~\cite{Wmap} imply that the standard cosmology is
given by the inflation and FRW universe~\cite{Inf}. Although there
exist a number of dark energy candidates, the best known
candidates are the cosmological constant  and the quintessence
scenarios. The equation of state (EOS) for the latter is
determined mostly  by the dynamics of  scalar and tachyon.

On the other hand, there exists the dynamical cosmological
constant derived  by the holographic principle.  Cohen {\it et al}
showed that in quantum field theory, the UV cutoff $\Lambda$ is
related to the IR cutoff $L_{\rm \Lambda}$ due to the limit set by
forming a black hole~\cite{CKN}. In other words, if $\rho_{\rm
\Lambda}$ is the quantum zero-point energy density caused by the
UV cutoff, the total energy of system with size $L_{\rm \Lambda}$
should not exceed the mass of the system-size black hole: $L_{\rm
\Lambda}^3 \rho_{\rm \Lambda}\le 2L_{\rm \Lambda}/G$. Here the
Newtonian constant  $G$ is related to the Planck mass
($G=1/M_p^2$). If the largest $L_{\rm \Lambda}$ is chosen to be
the one saturating this inequality,  the holographic energy
density is then given by $\rho_{\rm \Lambda}= 3c^2M_p^2/8\pi
L_{\rm \Lambda}^2$ with a factor $3c^2$. We consider $\rho_{\rm
\Lambda}$ as the dynamical cosmological constant.  Taking $L_{\rm
\Lambda}$ as the size of the present universe (Hubble horizon:
$R_{\rm HH}$), the resulting energy  is comparable to the present
dark energy~\cite{HMT}. Even though it leads to the data, this
 approach seems to be
incomplete. This is  because it fails to recover the equation of
state for a dark energy-dominated  universe~\cite{HSU}. In order
to resolve this situation, the two candidates for the IR cutoff
are proposed. One is the particle horizon $R_{\rm PH}$. This
provides $\rho_{\rm \Lambda} \sim a^{-2(1+1/c)}$ which gives the
equation of state $\omega_{\rm \Lambda}=1/3$ for $c=1$~\cite{LI}.
Unfortunately, it corresponds to  a decelerating universe. In
order to find an accelerating universe, one needs to introduce the
other known as the future event horizon $R_{\rm FH}$. In the case
of $L_{\rm \Lambda}=R_{\rm FH}$, one finds $\rho_{\rm \Lambda}
\sim a^{-2(1-1/c)}$ which may  describe the dark energy with
$\omega_{\rm \Lambda}=-1$ for $c=1$.  The related issues appeared
in Ref.\cite{FEH,Myung2}.

The above approach to dark energy have something to be clarified.
Usually, it is not an easy matter to determine the equation of
state for such a system with UV/IR cutoff. Actually, we have two
different views of  determining the equation of state for the
holographic energy density.  The first view is that its equation
of sates is not changing as the universe evolves\cite{HOV,WM}. It
is fixed by $p_{\rm \Lambda}=-\rho_{\rm \Lambda}$ initially. An
important point is that the holographic energy density  itself is
changing as a result of decaying into other matter. According to
the total energy-momentum conservation, its change must be
compensated by the corresponding change in other matter
sector\cite{Zim3}. In this case we need a source term to mediate
an interaction between two matters in the continuity
equations\cite{Zim1}. Here we note that the EOS for the
holographic energy and ordinary matter will be determined as the
same negative constant  by  the  interaction. As a result, two
matters are turned out to be imperfect fluids. We call this
picture as a decaying vacuum cosmology which may be  related to
the vacuum fluctuations\cite{PAD}.

The second view is that the equation of state for the holographic
energy density is not fixed but it is changing as the universe
expands without interaction\cite{HSU,LI}.  Even for being the
holographic matter $\rho_{\rm \Lambda}$ only, its equation of
state can be determined by the first Friedmann equation for
$L_{\rm \Lambda}=R_{\rm PH}$ and $R_{\rm FH}$. However, the
equation of state with $L_{\rm \Lambda}=R_{\rm HH}$ is not
determined by the first Friedmann equation\cite{HSU}.  It works
well for the presence of holographic and ordinary matter because
the energy-momentum conservation is required for each
matter\cite{LI}. Recently, it is shown that this picture works
even for the presence of interaction between the holographic
energy and ordinary matter\cite{WGA}.

In this work we study the role of  holographic energy density with
IR cutoff in the first view of the constant EOS. The key of our
system is an interaction between holographic energy and dust
matter. They is changing as a result of energy transfer from
holographic energy to dust matter.   The sourced Friedmann
equations are proposed  for a macroscopic system which can
describe the interaction between holographic and dust matter as
the universe expands.  We argue that an interaction induces an
acceleration.  Especially, we introduce the corresponding
microscopic model to provide cosmic anti-friction which  produces
an accelerating phase.

In the macroscopic picture,  we allow for an interaction between
holographic and ordinary matter. In general, the two continuity
equations are changed as\cite{Dan}

\begin{eqnarray}
\label{1eq1}&& \dot{\rho}_{\rm \Lambda}+3H \Big(\rho_{\rm
\Lambda}+p_{\rm \Lambda}\Big)=q_1,
~p_{\rm \Lambda}=\omega_{{\rm \Lambda}0} \rho_{\rm \Lambda},\\
\label{1eq2}&& \dot{\rho}_{\rm m}+3H(\rho_{\rm m}+p_{\rm m})=q_2,
~p_{\rm m}=\omega_{{\rm m}0} \rho_{\rm m}.
\end{eqnarray}
Here $H={\dot a}(t)/a(t)$ represents the Hubble parameter and the
overdots stand for  derivative with respect to the cosmic time
$t$. $\omega_{{\rm \Lambda}0}(\omega_{{\rm m}0})$ are the initial
EOS for the holographic energy (ordinary matter). The first
equation corresponds to the non-conservation of the holographic
matter, while the second represents the non-conservation of the
ordinary matter. Even though there exist  non-conservations, one
requires that the total energy-momentum be always conserved. The
second Friedmann equation is given by
\begin{equation}
\label{1eq3}\dot{H}=-\frac{4\pi}{M^2_p}\Big(\rho_{\rm
\Lambda}+p_{\rm \Lambda}+\rho_{\rm m}+p_{\rm m}\Big).
\end{equation}
 Integrating the
above  equation gives the sourced Friedmann equation
\begin{equation}
\label{1eq4} H^2=\frac{8\pi}{3M^2_p}\Big[
 \rho_{\rm \Lambda}+\rho_{\rm m}-\int^{t} q_1 dt-\int^{t} q_2 dt\Big].
\end{equation}
Usually the two Friedmann equations together with the continuity
equation are viewed on an equal footing so that only two of them
are independent. According to the thermodynamical approach to the
Friedmann equations~\cite{Dan}, the second Friedmann equation
(\ref{1eq3}) is more fundamental than the first Friedmann equation
(\ref{1eq4}). As a result, it is shown that Eq.(\ref{1eq3})
remains the same form even in the presence of sources but it is
not always true for Eq.(\ref{1eq4}). It is important to note that
both $\rho_{\rm \Lambda}$ and  $\rho_{\rm m}$ do not evolve
according to the $\omega_0$-parameters for their equations of
state because there exists an interaction between ordinary matter
and holographic matter. This  causes a continuous transfer of
energy from  holographic energy to ordinary matter/vice versa,
depending on the sign of two parameters $q_1$ and $q_2$.

\section{Microscopic mechanism for energy transfer}
 Let us imagine a universe made of cold dust matter with
$\omega_{{\rm m}0}=0$ but obeying the holographic principle. In
addition, suppose that the holographic energy density be allowed
to have any equation of state. But we here allow it to have
$\omega_{{\rm \Lambda}0}=-1$ for our purpose. If one chooses
$q_1=-q_2=-\Gamma\rho_{\rm m}$, their continuity equations take
the forms
\begin{eqnarray}
\label{2eq1}&& \dot{\rho}_{\rm \Lambda}=-\Gamma \rho_m, \\
\label{2eq2}&& \dot{\rho}_{\rm m}+3H\rho_{\rm m}=\Gamma \rho_m.
\end{eqnarray}
This means that the mutual interaction provides a decaying
process. That is, this is a decay of the holographic energy
component into pressureless matter with the decay rate $\Gamma$.
This process is necessarily accompanied by different equations
 of state  $\omega_{\rm m}$ and $\omega_{\rm \Lambda}$ even if they start with $\omega_{{\rm
m}0}=0$ and $\omega_{{\rm \Lambda}0}=-1$. The interaction induces
an accelerating expansion of the universe and determines their
equations of state solely. Actually, the accelerating phase arises
from  a largely effective non-equilibrium pressure
 $\Pi_{\rm m}$  defined as $\Pi_{\rm m}\equiv -\Gamma\rho_{\rm m}/3H(\Pi_{\rm
 \Lambda}=\Gamma\rho_{\rm m}/3H)$. Then  the two dynamic equations (\ref{2eq1})
and (\ref{2eq2}) are translated into two dissipative imperfect
fluids
\begin{eqnarray}
\label{2eq3}&& \dot{\rho}_{\rm \Lambda}+ 3H \Pi_{\rm
 \Lambda} =0, \\
\label{2eq4}&& \dot{\rho}_{\rm m}+3H(\rho_{\rm m}+\Pi_{\rm m})=0.
\end{eqnarray}
Now we introduce a microscopic mechanism to generate an
accelerating universe as a result of  energy transfer from the
holographic energy to pressureless matter. We recall that a
perfect fluid consists of particles with mass $m$ which move on
geodesics according to the geodesic equations: $m~dx^i/d\tau=p^i$
and $ Dp^i/d\tau=0$, where $\tau$ denotes the proper
time\cite{Zim2}. This corresponds to a Boltzmann equation for
one-particle distribution function $f(x,p)$

\begin{equation}
\label{2eq5}p^if_{,i}-\Gamma^{i}_{kl}p^kp^l \frac{\partial
f}{\partial p^i}=C[f].
\end{equation}
$C[f]$ is the Boltzmann collision integral which describes elastic
binary collisions between the particles. One of the second moments
for distribution function is the energy-momentum tensor for a
perfect fluid
\begin{equation}
\label{2eq6} T^{ik}=\int dP p^ip^k f(x,p)=\rho u^iu^k
+p(g^{ik}+u^iu^k)
\end{equation}
which satisfies the conservation law: $\dot{\rho}+ 3H(\rho+p)=0$.
Here $u^i$ is the macroscopic four-velocity.  Actually the
accelerating universe results from the cosmic anti-friction. An
anti-frictional force $F^i$, which is arisen from the surface
friction at interface between pressureless matter and holographic
energy, have exerted on the particles of the cosmic substratum. As
a result, the Boltzmann equation takes a different form with
$Dp^i/d\tau=F^i$
\begin{equation}
\label{2eq7}p^if_{,i}-\Gamma^{i}_{kl}p^kp^l \frac{\partial
f}{\partial p^i}+ m F^i \frac{\partial f}{\partial p^i}=C[f]
\end{equation}
which shows that the individual particle motion is no longer
geodesic. In a spatially homogeneous and isotropic universe, the
force $F^i$ may be  given by the general difference between the
macroscopic and particle velocities: $F^i=m(B u^i -Cu^i_{(p)})$,
where $B$ and $C$ are not constants but should depend on the
particle and fluid quantities, and $u^i_{(p)}$ is the particle
four-velocity.
 We achieve $B=C$ only for $u^i=u^i_{(p)}$ to guarantee that the mean
 motion remains force-free.
Under this condition, we have the microscopic form of force with
the particle energy $E=-p^iu_i$
\begin{equation}
\label{2eq8}m F^i=B(-Ep^i+m^2u^i)
\end{equation}
which makes the individual particles move on non-geodesic
trajectories, while the macroscopic mean motion remains geodesic
because for $p^i=mu^i$ and $ E=m$, $F^i$ vanishes consistently. In
the case that the cosmic substratum is non-relativistic dust
matter $(p_{\rm m} \ll \rho_{\rm m})$, the spatial projection of
force is reduced to
\begin{equation}
\label{2eq9}e_i F^i=-B m v,
\end{equation}
where $e^i=(p^i-Eu^i)/\sqrt{E^2-m^2}$ is the spatial direction of
the particle momentum and $v$ is the velocity of non-relativistic
particle. Eq.(\ref{2eq9}) is nothing but Stokes' law of friction.
For $B>0$, the force may be interpreted as cosmic friction  but
for $B<0$, as cosmic anti-friction. The quantity $B$ determines
the strength of the force and in turn the macroscopic interaction
between the holographic energy and dust matter. As was shown in
Ref.\cite{Zim2}, a microscopic force $F^i$-term in the Boltzmann
equation leads to the source term in the balances of second moment
of $f$ as
\begin{equation}
\label{2eq9-1}\dot{\rho}_{\rm m}+ 3H \rho_{\rm m}=-3B \rho_{\rm
m}.
\end{equation}
Comparing Eq.(\ref{2eq2}) with Eq.(\ref{2eq9-1}) leads to
$\Gamma=-3B$, which means that the (macroscopic) decay rate is
given by the (microscopic) cosmic-antifriction coefficient.
 With the definition of the imperfect pressure $\Pi_{\rm m}H=-\Gamma\rho_{\rm m}/3(=B\rho_{\rm
 m})$, the above energy balance becomes
\begin{equation} \label{2eq10}\dot{\rho}_{\rm m}+
3H (\rho_{\rm}+\Pi_{\rm m})=0.
\end{equation}
This proves that the action of force manifests itself as a
dissipative pressure macroscopically. In the next two section, we
provides two examples which determine the quantity $B$ by choosing
an explicit form of holographic energy density.

\section{Holographic energy density with the Hubble horizon}
In this section we choose the IR cutoff as Hubble horizon with
$L_{\rm \Lambda}=R_{\rm HH} \equiv 1/H$.
 Then the holographic dark energy  takes the form~\cite{LI}
\begin{equation} \label{2eq11}\rho_{\rm \Lambda}=\frac{3c^2  M^2_p H^2}{8\pi}.
\end{equation}
Here we consider the two interesting cases only. First we consider
the non-interacting case. In this case  the sourced Friedmann
equation (\ref{1eq4}) with $q_1=q_2=0$ can be simplified as
\begin{equation}
\label{2eq12}
 (1-c^2)H^2=\frac{8\pi}{3M^2_p}
 \rho_{\rm m}.
\end{equation}
For $c^2\not=1$, the first Friedmann equation takes the form of
$\rho_{\rm m}=3M_p^2(1-c^2)H^2/8\pi$. This implies that $\rho_{\rm
\Lambda}$ behaves as $\rho_{\rm m}$ because of $\rho_{\rm m} \sim
H^2 \sim \rho_{\rm \Lambda}$\cite{HSU,LI}. Choosing $\omega^{\rm
HH}_{\rm m}=0$ initially, one finds  a dust-like equation of state
for the holographic matter: $\omega^{\rm HH}_{\rm \Lambda}=0$.
This is not the case because the holographic energy density with
$\omega^{\rm HH}_{\rm m}=0$ cannot describe an accelerating
universe.

Hence  we study   an  interacting case of $q_1=-q_2 \equiv -q$
with $\omega^{\rm HH}_{{\rm \Lambda}0}=-1$ and $ \omega^{\rm
HH}_{{\rm m}0}=0$. This case was  discussed in the study of a
decaying vacuum cosmology\cite{HOV,Zim3,ENOW}. Here we have three
equations:
\begin{equation}
\label{2eq13} \dot{\rho}_{\rm \Lambda}=-q,~\dot{\rho}_{\rm
m}+3H\rho_{\rm m}=q,~ (1-c^2)H^2=\frac{8\pi}{3M^2_p}\rho_{\rm m}.
\end{equation}
In the case of $c^2\not=1$, differentiating  the last equation
with respect to the cosmic time and then using Eq.(\ref{1eq3})
leads to $\dot{\rho}_{\rm m}+3H\rho_{\rm m}=3c^2H\rho_{\rm m}$.
Comparing it with the second equation in Eq.(\ref{2eq13}), one
finds $q=\Gamma \rho_{\rm m}=3c^2H\rho_{\rm m}$. Their solution
are given by $\rho_{\rm m}=\rho_{{\rm
 m}0}a^{-3(1-c^2)}$ with the constant EOS $\omega^{\rm HH}_{\rm m}=-c^2$ and $\rho_{\rm \Lambda}
 =\frac{c^2 \rho_{{\rm m }0}}{1-c^2} a^{-3(1-c^2)}$ with the  EOS $\omega^{\rm HH}_{\rm
\Lambda}=-c^2$, respectively. Here $\rho_{{\rm
 m}0}$ is the initial value of $\rho_{\rm
 m}$. It is noted  that three equations in (\ref{2eq13}) are not
 enough to determine $c$ in the holographic energy density.
  Introducing the dissipative pressures
\begin{equation}
\label{2eq14} q=\Gamma\rho_{\rm m} \equiv -3H \Pi_{\rm m} \equiv
3H \Pi_{\rm \Lambda},
\end{equation}
one finds two imperfect fluids which satisfy
\begin{eqnarray}
\label{2eq15}&& \dot{\rho}_{\rm \Lambda}+ 3H \Pi_{\rm
 \Lambda} =0, \\
\label{2eq16}&& \dot{\rho}_{\rm m}+3H(\rho_{\rm m}+\Pi_{\rm m})=0.
\end{eqnarray}
In this case the quantity $B$ is determined as $B=-c^2H<0$. This
means that  the interaction between
 holographic energy  and dust  matter induces an accelerating
universe through the cosmic anti-friction if one chooses
$c^2>1/3$.  Unfortunately, the ordinary and holographic
 matter evolve in exactly the same way as
 $\omega_{\rm m}^{\rm HH}=\omega^{\rm HH}_{\rm \Lambda}=B/H=-c^2$.

\section{Holographic energy density with the future event horizon}

In order to find an accelerating universe which satisfies
\begin{equation} \label{4eq1}
\ddot{a}>0 \leftrightarrow \frac{d}{dt}\Big(\frac{1}{aH}\Big)<0
\leftrightarrow \omega<-\frac{1}{3},
\end{equation}
we need to take a shrinking comoving  Hubble scale, as was shown
in the inflationary universe. It means  that the changing rate of
$1/aH$ with respect to $a$ is always negative for an accelerating
universe. For this purpose, we introduce the future event horizon
$L_{\rm \Lambda}=R_{\rm FH} \equiv a \int_t^{\infty} (dt/a)=a
\int_a^{\infty}(da/Ha^2)$~\cite{LI,FEH}. In this case the sourced
Friedmann equation takes the form
\begin{equation}
\label{4eq2} H^2=\frac{8\pi}{3M^2_p}\Big[ \frac{3 c^2M^2_p}{8\pi
R^2_{\rm FH}} +\rho_{\rm m}-\int^{t} q_1 dt-\int^{t} q_2 dt\Big].
\end{equation}
Here we discuss the two interesting cases only. First we consider
the non-interacting case. In order to recover the known
non-interacting  solution, we consider the case that $\omega^{\rm
FH}_{\rm m}=0 $ for all time, while $\omega^{\rm FH}_{\rm
\Lambda}$ varies as the universe evolves. In this case we find the
effective EOS for the holographic energy density\cite{LI}
\begin{equation} \label{4eq3}
\omega^{\rm FH}_{\rm \Lambda}=-\frac{1}{3}-
\frac{2\sqrt{\Omega_{\rm \Lambda}}}{3c},
\end{equation}
where $\Omega_{\rm m}+\Omega_{\rm \Lambda}=1$ with $\Omega_{\rm
m}=8\pi \rho_{\rm m}/3M_p^2H^2$ and $\Omega_{\rm \Lambda}=8 \pi
\rho_{\rm \Lambda}/3M^2_pH^2$. This shows an accelerating phase.
Eq.(\ref{4eq3}) is a time-dependent EOS because $\Omega_{\rm
\Lambda}$ will be determined by solving the differential equation.
 Further, an interacting solutions for the presence
of ordinary and holographic matter appeared in\cite{WGA}

 Now we in a position to study the interacting case.  In this case we have three equations with
$\omega^{\rm HH}_{{\rm \Lambda}0}=-1, \omega^{\rm HH}_{{\rm
m}0}=0$:
\begin{equation} \label{4eq4}
\dot{\rho}_{\rm \Lambda}=-q, ~~\dot{\rho}_{\rm m}+3H\rho_{\rm
m}=q,~H^2=\frac{8\pi}{3M^2_p}\Big[
 \rho_{\rm \Lambda}+\rho_{\rm m}\Big].
\end{equation}
 With $q=\Gamma \rho_{\rm m}=-3B\rho_{\rm m} =\epsilon H \rho_{\rm m}$, the solution
to the above equations is given by
\begin{equation} \label{4eq5}
\rho_{\rm m} = \rho_{{\rm m}0}a^{-3(1-\epsilon/3)},~\rho_{\rm
\Lambda} =\frac{\epsilon \rho_{{\rm m}0}}{3-\epsilon}
a^{-3(1-\epsilon/3)}, ~\epsilon=1+ \frac{2}{3c^2}\pm
\frac{2\sqrt{3c^2+1}}{3c^2}.
\end{equation}
We observe that due to the interaction, the ordinary matter no
longer scales like $\rho_{\rm m} \sim a^{-3}$. In the case of
$c^2=1$, one has $\epsilon=1/3(-)$ and $\epsilon=3(+)$. The first
case corresponds to a decelerating universe which contains a
reduced form of dust-matter with $\rho_{\rm m}(\rho_{\rm \Lambda}
)\sim a^{-8/3}$, while the last case is an accelerating universe
with the cosmological constant $\rho_{\rm m}(\rho_{\rm \Lambda})
\sim {\rm const}$. We note that the case of $\epsilon=1/3$
corresponds to the particle horizon even though it is derived from
the future event horizon.

However, the ordinary and holographic matter evolve in exactly the
same way. Hence we may confront with the same trouble as other
$\Lambda(t)$ CDM cosmology\cite{WM}. Here the anti-friction
coefficient $B $ is given by $B/H=\omega^{\rm FH}_{\rm
m}=-\frac{\epsilon}{3}$. Finally, the constant EOS for the future
event horizon  is given by
\begin{equation} \label{4eq6}\omega^{\rm FH}_{\rm m}=\omega^{\rm FH}_{\rm \Lambda}=-\frac{1}{3}
-\frac{2}{9c^2}-\frac{2\sqrt{3c^2+1}}{9c^2}.
\end{equation}

\section{Discussions}

 Introducing an interaction between
the holographic energy density and dust matter, we obtain the
enhanced information on the equation of state $\omega^{\rm
HH}_{\rm \Lambda}=-c^2$ for the holographic energy density
$\rho_{\rm \Lambda}=3c^2M_p^2/8\pi L_{\rm \Lambda}^2$ with $L_{\rm
\Lambda}=1/H$. Without interaction, One finds that the equation of
state is fixed to be $\omega^{\rm HH}_{\rm \Lambda}=0$ as that for
a dust matter. However, the ordinary and holographic
 matter evolve in exactly the same way. This may induce a trouble of
 the indifference between the holographic energy density and
 ordinary matter in the future universe. In the case of the
 interaction between
 holographic energy density with the future event horizon and dust matter,
  we find the similiar case but with the
 different EOS as is shown in Eq.(\ref{4eq6}).

It is not an easy matter to determine the equation of state for a
system with UV or IR cutoff.
 As  another example, we introduce the
perturbations of inflation in the early universe. In the
transplanckian approach to inflation with UV cutoff $\Lambda$, the
equation of state for quantum and classical fluctuations of
inflation depends on the scheme of a calculation. It is usually
given by $\omega_{\rm qfi/cfi}=1/3$ without the transplanckian
backreaction. Even for
 quantum fluctuations of inflation, a constant energy
density $\rho_{\rm qfi} \sim (\Lambda H)^2$ with $\omega_{\rm
qfi}=1/3,\dot{H}=0$ is not compatible with the continuity
equation: $\dot{\rho}_{\rm qfi}+3H \rho_{\rm qfi}(1+\omega_{\rm
qfi})=0$\cite{KS}. To restore the compatibility, the continuity
equation should be modified as $\dot{\rho}_{\rm qfi}+3H \rho_{\rm
qfi}(1+\omega_{\rm qfi})=q$ with a source $q$. Then this modified
equation is satisfied with $q=4H\rho_{\rm qfi}$. Furthermore, it
gives $\omega_{\rm qfi/cfi}\simeq -1$ when including the
transplanckian backreaction with a non-linear dispersion
relation~\cite{BM}. On the other hand, if one includes the effects
of transplanckian backreaction through the sourced Friedmann
equations~\cite{Dan}, it provides a different result of
$\omega_{\rm qfi}\simeq 1/3(1-4\Lambda^2/M_p^2)$.

Consequently,  our system is composed of  holographic energy and
dust matter with their interaction $\Gamma \rho_{\rm m}$. They are
changing as a result of decaying from holographic energy into dust
matter with decay rate $\Gamma$. The sourced Friedmann equations
are suitable for a macroscopic system which can describe the
interaction between holographic and dust matter as the universe
expands. For clarity, we introduce the corresponding microscopic
model to provide cosmic anti-friction
 $B=-3/\Gamma$ which plays a key role in producing an accelerating phase.
Finally, we argue that an interaction induces an acceleration.

\section*{Acknowledgment}
I thank P. Wang for helpful discussions. This work  was in part
supported by the SRC Program of the KOSEF through the Center for
Quantum Spacetime (CQUeST) of Sogang University with grant number
R11-2005-021-03001-0.


\begin{thebibliography}{99}
\bibitem{SN} S. J. Perllmutter {\it et al.}, Astrophys. J. {\bf 517},
565(1999)[astro-ph/9812133
]; 
A. G. Reiss {\it et al.}, Astron. J. {\bf 116},
1009 (1998)[astro-ph/9805201 ]; 
A. G. Reiss  {\it et al.}, Astrophys. J. {\bf 607},
665(2004)[astro-ph/0402512]; 

\bibitem{Wmap} H. V. Peiris  {\it et al.}, Astrophys. J. Suppl. {\bf 148} (2003) 213
[astro-ph/0302225];
C. L. Bennett  {\it et al.}, Astrophys. J. Suppl. {\bf 148} (2003)
1[astro-ph/0302207];
D. N. Spergel  {\it et al.}, Astrophys. J. Suppl. {\bf 148} (2003)
175[astro-ph/0302209].

\bibitem{Inf} A.~H. ~Guth,
Phys.\ Rev.\ D {\bf 23}, 347 (1981);
A.~D.~Linde, Phys.\ Lett.\ B {\bf 108}, 389 (1982);
A.~Albrecht and P.~J.~Steinhardt, Phys.\ Rev.\ Lett.\  {\bf 48},
1220 (1982).




\bibitem{CKN} A. Cohen, D. Kaplan, and A. Nelson, Phys. Rev. Lett.
{\bf 82}, 4971 (1999)[arXiv:hep-th/9803132].

\bibitem{HMT} P. Horava and D. Minic, Phys. Rev. Lett.
{\bf85}, 1610 (2000)[arXiv:hep-th/0001145];
S. Thomas, Phys. Rev. Lett. {\bf 89}, 081301 (2002).



\bibitem{HSU} S. D. Hsu, Phys. Lett. B {\bf 594},13
(2004)[hep-th/0403052].



\bibitem{LI} M. Li, Phys. Lett. B {\bf 603}, 1
(2004)[hep-th/0403127].

\bibitem{FEH}
Q-C. Huang and Y. Gong, JCAP {\bf 0408}, 006
(2004)[astro-ph/0403590];
Y. Gong, Phys. Rev. D {\bf 70}, 064029 (2004)[hep-th/0404030];
B. Wang, E. Abdalla and Ru-Keng Su, hep-th/0404057;
K. Enqvist and M. S. Sloth, Phys. Rev. Lett. {\bf 93}, 221302
(2004) [hep-th/0406019];
S. Hsu and  A. Zee, hep-th/0406142;
K. Ke and  M. Li, hep-th/0407056;
P. F. Gonzalez-Diaz, hep-th/0411070;
S. Nobbenhuis, gr-qc/0411093;
H. Kim, H. W. Lee, and Y. S. Myung, hep-th/0501118;
gr-qc/0507010.

\bibitem{Myung2} Y. S. Myung, Phys. Lett. B {\bf 610}, 18 (2005)[hep-th/0412224];
Mod. Phys. Lett. A {\bf 27}, 2035 (2005) [hep-th/0501023];
A. J. M. Medved, hep-th/0501100.



\bibitem{HOV} R. Horvat, Phys. Rev. D {\bf 70}, 087301 (2004)
[astro-ph/0404204].

\bibitem{WM} P. Wang and X. Meng, Class. Quant. Grav. {\bf 22}, 283(2005)
[astro-ph/0408495].

\bibitem{Zim3}  D.  Pavon and W. Zimdahl, gr-qc/0505020;
W. Zimdahl, gr-qc/0505056;

\bibitem{Zim1} W. Zimdahl, D.  Pavon, and L. P. Cimento, Phys. Lett. B {\bf 521}, 133 (2001)
[astro-ph/0105479].



\bibitem{PAD} T. Padmanabhan, astro-ph/0411044.

\bibitem{WGA} B. Wang, Y. Gong, and E. Abdalla, hepth/0505069.

\bibitem{Dan}
U. H. Danielsson, Phys. Rev. D {\bf 71}, 023516 (2005)
[hep-th/0411172];

\bibitem{Zim2} W. Zimdahl, D. J. Schwarz, A. B. Balakin, and D.
Pavon, Phys. Rev. D {\bf 64}, 063501 (2001) [astro-ph/0009353];

A. B. Balakin, D. Pavon,  D. J. Schwarz, and W. Zimdahl,  New J.
Phys. {\bf 5}, 085 (2003) [astro-ph/0302150].



\bibitem{ENOW} E. Elizalde, S. Nojiri, S. D. Odintsov, and P.
Wang, hep-th/0502082.

\bibitem{KS} E. Keski-Vakkuri and M. S. Sloth, JCAP {\bf 0308}, 001
(2003) [hep-th/0306070].

\bibitem{BM} R. H. Brandenberger and  J. Martin, Phys. Rev. {\bf D}71, 023504 (2005)
[hep-th/0410223];





\end{thebibliography}
\end{document}